\documentclass[
    ,final            % use final for the camera ready runs
  ]
  {aipproc}

\layoutstyle{8x11single}

\newcommand{\bea}{\begin{eqnarray}}    
\newcommand{\eea}{\end{eqnarray}}      
\newcommand{\be}{\begin{equation}}
\newcommand{\ee}{\end{equation}}
\newcommand{\bef}{\begin{figue}}
\newcommand{\eef}{\end{figure}}
\begin{document}

\title{Characterizing the large scale inhomogeneity of  the galaxy distribution
\footnote{In the proceedings of the ``Invisible Universe
International Conference'', AIP proceedings series.}}


\author{Francesco Sylos Labini}{
  address={Museo Storico della Fisica e Centro Studi e Ricerche Enrico
  Fermi, - Piazzale del Viminale 1, 00184 Rome, Italy \& Istituto dei
  Sistemi Complessi CNR, - Via dei Taurini 19, 00185 Rome, Italy } }

\begin{abstract}
In order to investigate whether galaxy structures are compatible with
the predictions of the standard LCDM cosmology, we focus here on the
analysis of several simple and basic statistical properties of the
galaxy density field. Namely, we test whether, on large enough scales
(i.e., $r>10$ Mpc/h), this is self-averaging, uniform and
characterized by a Gaussian probability density function of
fluctuations. These are three different and clear predictions of the
LCDM cosmology which are fulfilled in mock galaxy catalogs generated
from cosmological N-body simulations representing this model.  We
consider some simple statistical measurements able to tests these
properties in a finite sample. We discuss that the analysis of several
samples of the Two Degree Field Galaxy Redshift Survey and of the
Sloan Digital Sky Survey show that galaxy structures are non
self-averaging and inhomogeneous on scales of $\sim$ 100 Mpc/h, and
are thus intrinsically different from LCDM model predictions.
Correspondingly the probability density function of fluctuations shows
a "fat tail" and it is thus different from the Gaussian prediction.
Finally we discuss other recent observations which are odds with LCDM
predictions and which are, at least theoretically, compatible with the
highly inhomogeneous nature of galaxy distribution. We point out that
inhomogeneous structures can be fully compatible with statistical
isotropy and homogeneity, and thus with a relaxed version of the
Cosmological Principle.
\end{abstract}

\maketitle

%%%%%%%%%%%%%%%%%%%%%%%%%%%%%%%%%%%%%%%%%%%%
%% MAINMATTER
%%%%%%%%%%%%%%%%%%%%%%%%%%%%%%%%%%%%%%%%%%%%

\section{Introduction}

In the past twenty years many observations have been dedicated to the
study of the large scale distributions of galaxies
\cite{cfa1,cfa2,pp,ssrs2,lcrs,sdss,colless01}.  In particular during
the last decade two ambitious observational programs have measured the
redshift of more than one million objects \cite{sdss,colless01}.  All
these surveys have detected larger and larger structures, thus finding
that galaxies are organized in a complex network of clusters,
super-clusters, filaments and voids. For instance the famous ``slice
of the Universe'', that represented the first set of observations done
for the CfA Redshift Survey in 1985 \cite{dlhg85}, mapped
spectroscopic observations of about 1100 galaxies in a strip on the
sky 6 degrees wide and about 130 degrees long. This initial map was
quite surprising, showing that the distribution of galaxies in space
was anything but random, with galaxies actually appearing to be
distributed on surfaces, almost bubble like, surrounding large empty
regions, or ``voids.''.  The structure running all the way across the
survey between 50 and 100 Mpc/h\footnote{We use $H_0=100 h$ km/sec/Mpc
  for the value of the Hubble constant.} was called the ``Great Wall''
and at the time of the discovery was the largest single structure
detected in any redshift survey. Its dimensions, limited only by the
sample size, are about $200\times 80 \times 10 $ Mpc/h, a sort of like
a giant quilt of galaxies across the sky \cite{gh89}. More and more
galaxy large scale structures were identified in the other redshift
surveys such as the Perseus-Pisces super-cluster \cite{pp} which is
one of two dominant concentrations of galaxies in the nearby
universe. This long chain of galaxies lies next to the the so-called
Taurus void, which is a large circular void bounded by walls of
galaxies on either side of it. The void has a diameter of about 30
Mpc/h.  Few years ago, in the larger sample provided by the Sloan
Digital Sky Survey (SDSS), it has been discovered the Sloan Great Wall
\cite{sloangreatwall}, which is a giant wall of galaxies which may be
the largest known structure in the Universe, being nearly three times
longer than the Great Wall. The discovery of larger and larger
structures was surprising because standard cosmological models
unambiguously predict that fluctuations should be small on large
scales with rapidly decaying correlations; i.e., the spatial extension
of structures in these models should be limited to some tens Mpc/h.

\section{Predictions of LCDM models}

\begin{figure}
 \includegraphics*[height=.5\textheight]{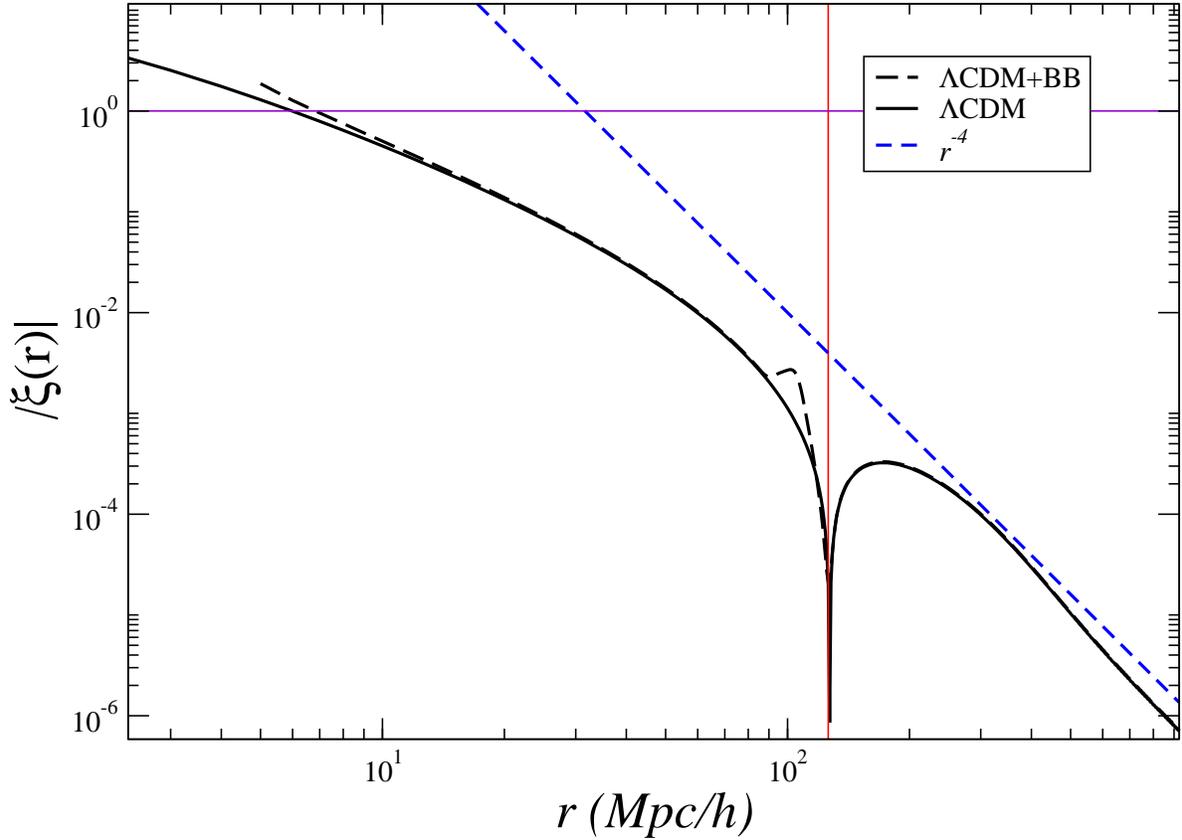} 
\caption{Absolute
  value of the two-point correlation function for the $\Lambda$CDM
  model and for the same model with the baryonic bump (BB) at $\sim $
  100 Mpc/h (adapted from \cite{cdm_theo}). A reference line with
  power law index $-4$ is reported to illustrate the behavior for
  $r>150$ Mpc/h where the correlation function becomes negative. }
\label{fig1} 
\end{figure}

Before discussing  whether the large scale structures identified in
galaxy catalogs are {\it compatible} with the prediction of the
standard LCDM cosmology, let us briefly discuss which, are the main
features of the galaxy two-point correlation function $\xi(r)$
according to this model (see Fig.\ref{fig1}). There are three
different regimes and three different length scales $r_0, r_c$ and
$r_{bao}$ \cite{glass,cdm_theo,bao}.  For $r>r_0$ the behavior is fixed by
the physics of the early universe, being thus an imprint of the
initial conditions.
(i) On scales smaller than $r_0$, where $\xi(r_0)=1$, matter
distribution is characterized by strong clustering; 
i.e. $\xi(r) \gg
  1$, 
about which little is known analytically and which is generally
constrained by N-body simulations where it is typically found that,
for $r<r_0$, $\xi(r) \sim r^{-\gamma}$ with $\gamma \approx 1.5$
(ii) The second length scale is such that $\xi(r_c) =0$, and it is
located at $r_c \gg r_0$ \cite{glass}.
In the range of scales $r_0 < r < r_c$, $\xi(r)$ is characterized by
small-amplitude positive correlations, which rapidly decay to zero
when $r \rightarrow r_c$.  The third length scale $r_{bao}$ is located
on scales of the order of  $r_c$. This is the
real-space scale corresponding to the baryon acoustic oscillations
(BAO) at the recombination epoch.  
(iii) Finally in the third range of scales, namely for $r>r_c$,
$\xi(r)$ is characterized by a negative power-law behavior,
i.e. $\xi(r) \sim - r^{-4}$ \cite{glass,book}.  Positive and negative
correlations are exactly balanced in such a way that
$\int_0^{\infty} \xi(r) d^3r =0$. This is a global condition on the
system fluctuations, which corresponds to the super-homogeneous nature
of the matter distribution \cite{glass,book}; i.e., that this
characterized by a sort of stochastic order and by fluctuations that
are depressed with respect to a purely uncorrelated distribution of
matter (i.e. white noise). This corresponds to the linear behavior of
the matter power spectrum as a function of the wave-number $k$ for
$k\rightarrow 0$ (named the Harrison-Zeldovich tail), and it
characterizes not only the LCDM model but all models of density
fluctuations in the framework of the Friedmann-Robertson-Walker metric
\cite{glass,book}. Roughly, 
structures correspond to positive correlations, while negative
correlations correspond to under-densities; thus the length-scale
$r_c$ can be regarded as an upper limit to the spatial extension of
structures in these models.

To summarize: on large enough scales $r>r_0\approx 10$ Mpc/h the
predicted density field must be uniform and weakly correlated (and
thus self-averaging --- see below). In these models, on large enough
scales, fluctuations are small and thus gravitational clustering in an
expanding universe gives a simple prediction, as in the linear regime
initial fluctuations are only linearly amplified during the
growth \cite{bias,cdm_theo}. Correspondingly the probability density
function (PDF) of fluctuations remains substantially Gaussian. Indeed,
the central limit theorem can be broken when correlations are
long-ranged and strong \cite{book}, while these models predict weak
positive correlations beyond 10 Mpc/h and up to 150 Mpc/h and very
weak negative correlations for $r>150$ Mpc/h.

%%%%%%%%%%%%%%%%%%%%%%%%%

\section{Recent results on galaxy correlations}

There is an extensive literature on the measurement galaxy
correlations properties; here we briefly review some recent results
which are useful to point out the status of the art in the field (we
refer the interested reader
to \cite{2df_epl,2df_aea,sdss_epl,sdss_aea,bao} for more details).
Few years ago Eisenstein et al. \cite{lrg} determined the galaxy
two-point correlation function in a preliminary luminous red galaxy
(LRG) sample of the Sloan Digital Sky Survey (SDSS) data release 3
(DR3). They found that $r_0 \approx 15$ Mpc/h, $r_{bao} \approx 110$
Mpc/h and $r_c \approx 150$ Mpc/h, thus claiming for an overall
agreement with the LCDM prediction and for a positive detection of the
scale $r_{bao}$ at about 110 Mpc/h. More recently other
authors \cite{cabre,martinez,kazin} measured the same estimator of the
correlation function in the LRG-DR6 sample and in the LRG-DR7
sample. They found that $r_c \approx 200$ Mpc/h and that
$r_{boa} \approx 100$ Mpc/h, although they also showed, by considering
many realizations of mock galaxy catalogs, that the expected variance
of the model is large enough that the observed baryonic acoustic peak
and larger scale signal are {\it only} consistent with LCDM at the
1.5$\sigma$ level. In addition Martinez et al. \cite{martinez}
measured that $r_c\approx 50 $ Mpc/h in the Two Degree Field Galaxy
Redshift Survey (2dFGRS) samples; they claimed that $r_{bao}$ is
detectable when the correlation function is negative. However this is
not what one expects in the context of the LCDM model where, as
discussed above, the zero point of the correlation function must be a
single scale for any type of objects. Indeed both linear gravitational
clustering, and biasing (i.e., threshold selection) the density field
give rise to a linear amplification of the amplitude of $\xi(r)$ 
\cite{cdm_theo,bao}.

In summary different authors
found, in different samples, different values for the three scales
$r_0,r_c,r_{bao}$. These results are puzzling as the model gives an
unambiguous prediction for the scales $r_c$ and $r_{bao}$, leaving
however undetermined, but in the range [5,15] Mpc/h, the scale
$r_0$. This latter behavior is generally ascribed to a luminosity
selection effect --- luminosity bias.  A little discussion is deserved
to the problem of bias and variance in estimators, which can instead
be a major issue in these analyses
\cite{cdm_theo,bao}.

There have been published other very puzzling results which seem to be
in contrast with the above mentioned results. For instance,
Loveday \cite{loveday}, by measuring the redshift-dependent luminosity
function and the comoving radial density of galaxies in the SDSS-DR1,
found that the apparent number density of bright galaxies increases by
a factor $\approx $ 3 as redshift increases from $z = 0$ to $z =
0.3$ \cite{loveday}.  To explain these observations a significant
evolution in the luminosity and/or number density of galaxies at
redshifts $z < 0.3$ has then been proposed \cite{loveday}. However an
independent test has not been provided to support such a conclusion.
Independently of the origin of this density growth, it remains that
major issue that the density is not constant and thus its
determination inside this sample is very ambiguous. As the correlation
function $\xi(r)$ measures the amplitude and the scale dependence of
correlations between density fluctuations {\it normalized} to the
sample value of the density, if this quantity is not well-defined and
determined with a small error, there can be substantial systematic
(i.e. volume-dependent) effects on its estimation.

Another puzzling observation is represented by a CCD survey of bright
galaxies within the Northern and Southern strips of
2dFGRS \cite{colless01}. This shows conclusive evidences of
fluctuations of $ \sim 30\%$ in galaxy counts as a function
of apparent magnitude \cite{busswell03} (see also
\cite{frith03,frith06} for similar observations in other galaxy
samples).  Further since in the angular region toward the Southern
galactic cap a deficiency, with respect to the Northern galactic cap,
in the counts below magnitude $\sim 17$ was found, persisting over the
full area of the APM and APMBGC catalogs, this would be an evidence
that there is a large void of radius of about $150$ Mpc/h implying
that there are spatial correlations extending to scales larger than
the scale detected by the 2dFGRS correlation
function \cite{norbergxi01,norbergxi02}. Indeed, by considering the
two-point correlation function, and thus by normalizing the amplitude
of fluctuations to the estimation of the sample density, the
length-scale $r_0 \approx 6-8$ Mpc/h was derived
\cite{norbergxi01,norbergxi02}. 
Structures and fluctuations at scales of the order of 100 Mpc/h or
more are at odds with the prediction of the concordance model of
galaxy formation
\cite{busswell03,frith03,frith06}, while
the small value of the correlation length is indeed compatible.

Summarizing the current observational situation: there are evidences
that galaxy distribution exhibits large amplitude fluctuations on
scales of $\sim 100$ Mpc/h. On the other hand there are many results
showing that $r_0 \approx 10$ Mpc/h, while there is a substantial
indetermination on the value of $r_c$, although in the very large
volumes of the LRG samples it was found $r_c \approx 200$ Mpc/h.  In
what follows we try to clarify this puzzling situation, i.e. the
coexistence of the small typical length scale $r_0$ measured by the
two-point correlation function analysis with the large fluctuations in
the galaxy density field on large scales as measured by the simple
galaxy counts.  The problem is that because of the large fluctuations
in galaxy counts, the estimation of the sample density is not stable
and thus one must critically consider the significance of the
normalization of fluctuations amplitude to the estimation of the
sample density as used in the correlation analysis employed to measure
the length scales $r_0,r_c$ and $r_{bao}$.

\section{Testing theoretical models against data} 

 In evaluating whether a model (CDM) is consistent with the data, it
 should be shown that {\it at least the main} statistical properties
 of the model are indeed consistent with the data. There is a number
 of different properties which can one consider and which are useful
 to test the assumptions that have been used when one studies the
 consistence of a theoretical model with real data by only considering 
 the behavior of $\xi(r)$. Namely the assumptions of (i) self-averaging
 (ii) uniformity (or spatial homogeneity).  When, inside the given
 sample, the assumption (i) and/or (ii) are/is violated then the
 conclusion is that CDM model is not compatible with the properties of
 the data \cite{2df_epl,2df_aea,sdss_aea,sdss_epl}.  In general little
 attention is deserved to these properties and one generally assumed
 that these are satisfied inside a given sample focusing directly on
 the behavior of $\xi(r)$ or that of its Fourier conjugate, the power
 spectrum. Indeed, the use of any statistical quantity which is
 normalized to some estimations of the sample density implicitly
 assume that a distribution, inside the given finite sample, is
 self-averaging and uniform.

\begin{figure}
\includegraphics[height=.3\textheight]{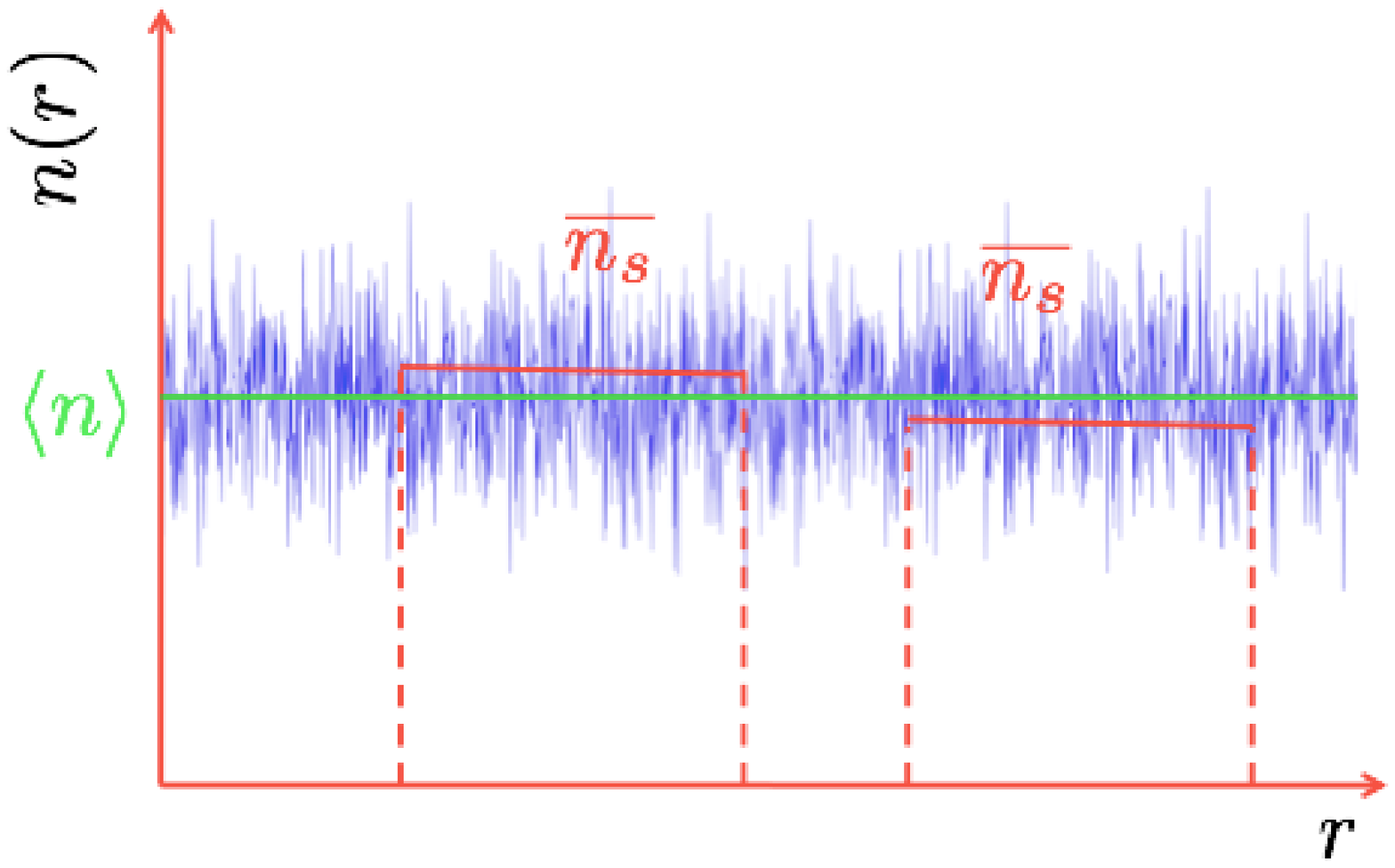}      
\includegraphics[height=.3\textheight]{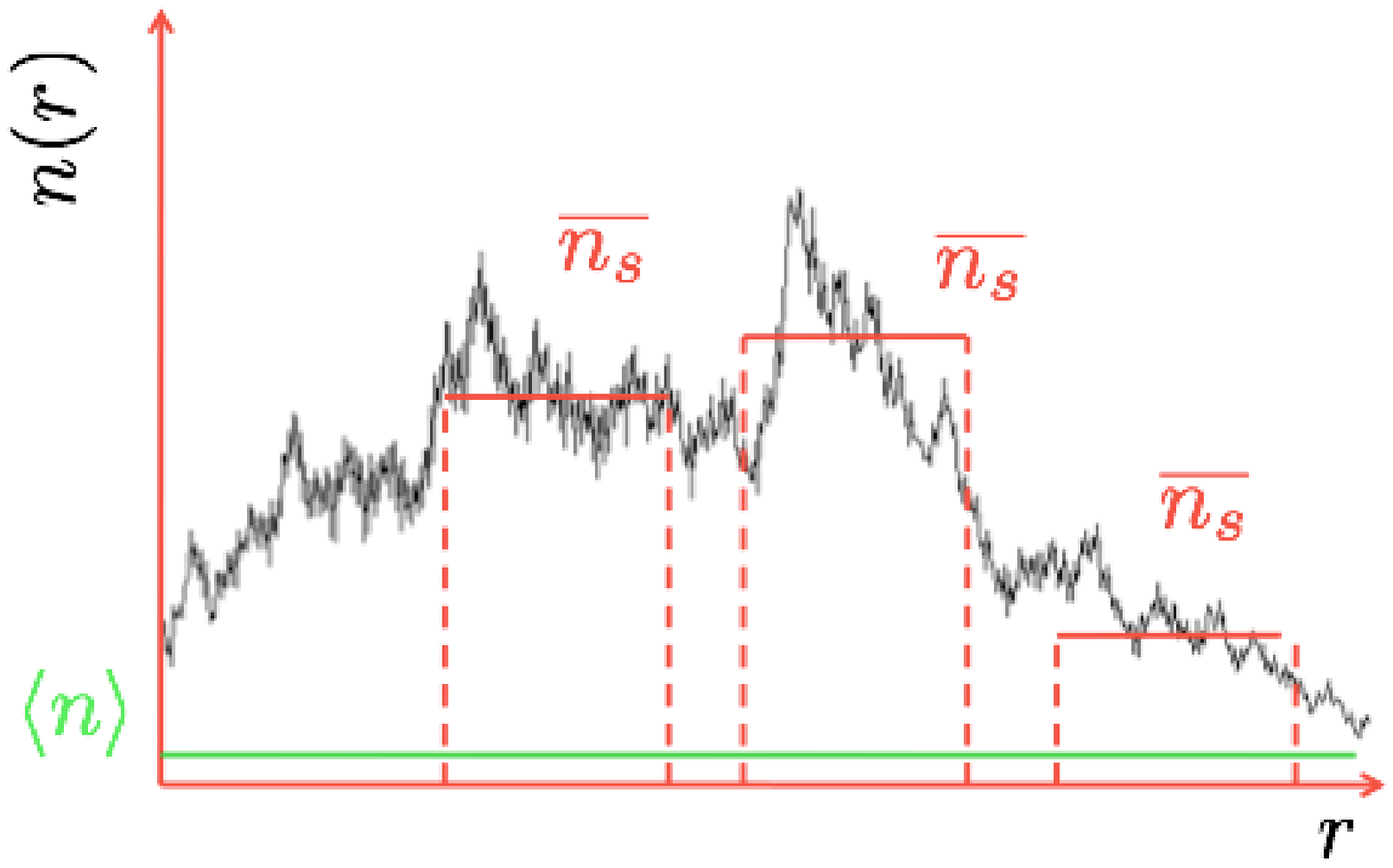}   
\caption{Left panel: One
  dimensional density field with small amplitude fluctuations.  Right
panel: One dimensional density field with large amplitude
fluctuations}
\label{fig2}
\end{figure}

 In order to illustrate the problems underlying these assumptions, let
 us consider a simple one-dimensional density field. In Fig.\ref{fig2}
 (left panel)  it is shown the case of one dimensional density field
 with small amplitude fluctuations. In red we report two examples of
 samples with finite spatial extension.  In each of the two cases one
 measures the sample density $\overline{n}$ which is ``close'' to the
 asymptotic average density $\langle n \rangle$ (i.e., the ensemble
 average density or the average density measured in the infinite
 volume limit). The rate of the difference between $\overline{n}$ and
 $\langle n \rangle$ with scales is determined by the behavior of the
 two-point correlation function \cite{book}.  A completely different
 situation is represented by the one-dimensional density field shown
 in Fig.\ref{fig2} (right panel). In this case the field is
 characterized by large fluctuations and the determination of the
 sample density $\overline{n}$ in different regions (red lines) does
 not give a useful information about $\langle n \rangle$.

 It is then clear that depending on the properties of the density
field, one should adopt a certain statistical characterization or
another. While in the case shown in the left panel of Fig.\ref{fig2}
it is meaningful to normalize fluctuations to the sample estimate of
the density, this becomes meaningless in the case shown in the right
panel of Fig.\ref{fig2}. Thus, in a {\it finite sample} one needs to
set up a strategy to test the different assumptions used in the
statistical analysis. To this aim one has to make a clear distinction
between statistical quantities which are normalized to the sample
average density and those which are not. If one wish to determine
whether a statistically meaningful estimate of the average density is
possible in the given samples, one should use statistical quantities
that do not require the assumption of homogeneity inside the sample
and thus avoid the normalization of fluctuations to the estimation of
the sample average. These are thus conditional quantities, as for
example the conditional density $n_i(r)$ from the $i^{th}$ galaxy,
which gives the density in a sphere of radius $r$ centered on the
$i^{th}$ galaxy. Conditional quantities are well-defined both in the
case of homogeneous and inhomogeneous point distributions, as they
require only local (i.e., not global) determination of statistical
properties \cite{sdss_aea}.

\begin{figure}
\includegraphics*[height=.4\textheight]{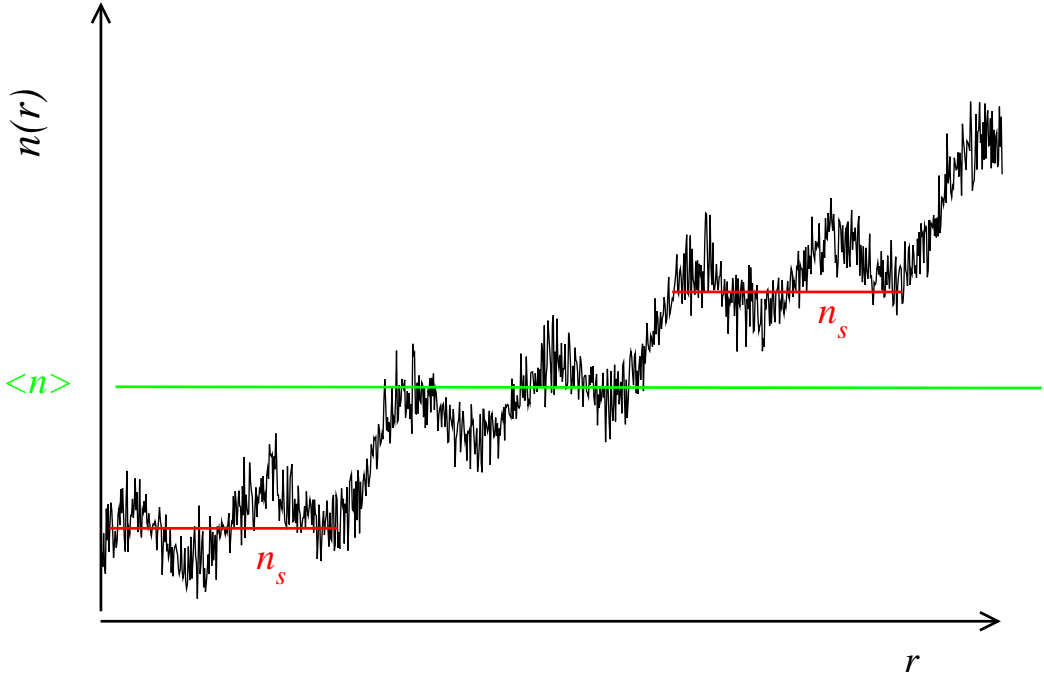}      
\caption{One
  dimensional density field with large fluctuations: this density
field is not self-averaging inside the considered sample, as the local
density varies in a systematic way. }
\label{fig2b}
\end{figure}

In general one should consider also another important complication.
Statistical properties are determined by making averages over the
whole sample volume~\cite{book}.  In doing so, one implicitly assumes
that a certain quantity measured in different regions of the sample is
statistically stable, i.e., that fluctuations in different sub-regions
are described by the same probability density function (PDF).  However
it may happen that measurements in different sub-regions show
systematic (i.e., not statistical) differences, which depend, for
instance, on the spatial position of the specific sub-regions (see
Fig.\ref{fig2b}).
In
this case the considered statistic is not statistically stationary in
space, the PDF systematically differs in different sub-regions and its
whole-sample average value is not a meaningful descriptor \cite{book}.
In general such systematic differences may be related to two different
possibilities: (i) that the underlying distribution is not
translationally and/or rotationally invariant; (ii) that the volumes
considered are not large enough for fluctuations to be
self-averaging \cite{sdss_epl,sdss_aea}.

For this reason, our first aim is to study whether galaxy distribution
  is self-averaging by characterizing conditional fluctuations. If the
  distribution is self-averaging, then one can consider whole-sample
  conditional average quantities and study the possible transition
  from non-uniformity to uniformity. This can be achieved by
  characterizing the behavior of, for instance, the conditional
  density. If the distribution is uniform, or becomes uniform at a
  certain scale smaller than the sample size, one can characterize the
  (residual) correlations between density fluctuations by studying the
  standard two-point correlation function. Therefore the consideration
  of $\xi(r)$ is the last point in this list, and it is appropriate
  only if one has proved that the distribution is self-averaging and
  uniform inside the given sample. We make similar tests in the real
  samples and in the mock catalogs which represent the model
  predictions.  In this way we test whether large scale structures
  identified in galaxy catalogs are {\it compatible} with the
  prediction of the standard LCDM cosmology,

\section{Statistical properties of 
galaxy distribution in the 2dFGRS and SDSS samples}

The main stochastic variable which we consider and of which we
determine statistical properties is the conditional number of points
in spheres mentioned above.  Namely we compute for each scale $r$ the
$\{ N_i(r) \}_{i=1...M}$ determinations of the number of points inside
a sphere of radius $r$ centered on the $i^{th}$
galaxy \cite{sdss_aea}.  The random variable $N_i(r)$ depends thus on
the scale $r$ and on the spatial position of the sphere's center; we
can express the $i^{th}$ sphere center coordinates with its radial
distance $R_i$ and with its angular coordinates $\vec{\alpha}_i =
(\eta_i,\lambda_i)$. Thus, in general, we can write%
\be 
\label{eq1a}
N_i(r) = N(r; R_i, \vec{\alpha}_i) \;. 
\ee
 When we integrate over the angular coordinates $\vec{\alpha}_i$ for
  fixed radial distance $R_i$ we have that $N_i(r)=N(r; R_i)$, i.e. it
  depends on two variables the length-scale of the sphere $r$ and the
  distance-scale of the $i^{th}$ sphere center $R_i$ and thus it has
  been called the scale-length analysis \cite{sdss_aea,sdss_epl}.

\begin{figure}
  \includegraphics[height=.5\textheight]{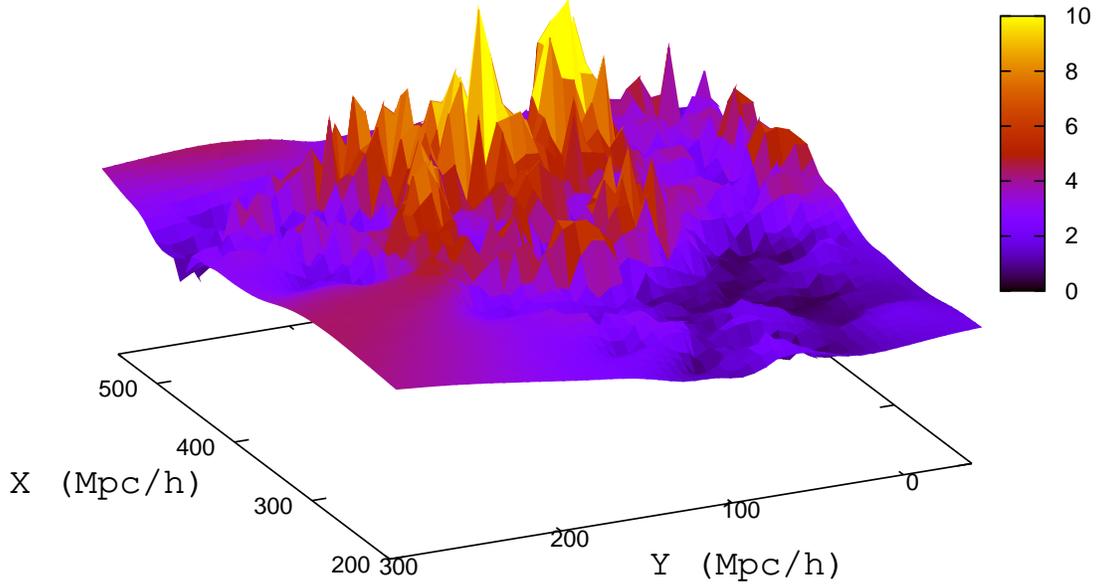} \caption{Three
  dimensional representation of the SL analysis with $r=10$ Mpc/h in a
  SDSS sample (adapted from \cite{sdss_aea}). }
\label{fig4} 
\end{figure}

\begin{figure}
  \includegraphics[height=.5\textheight]{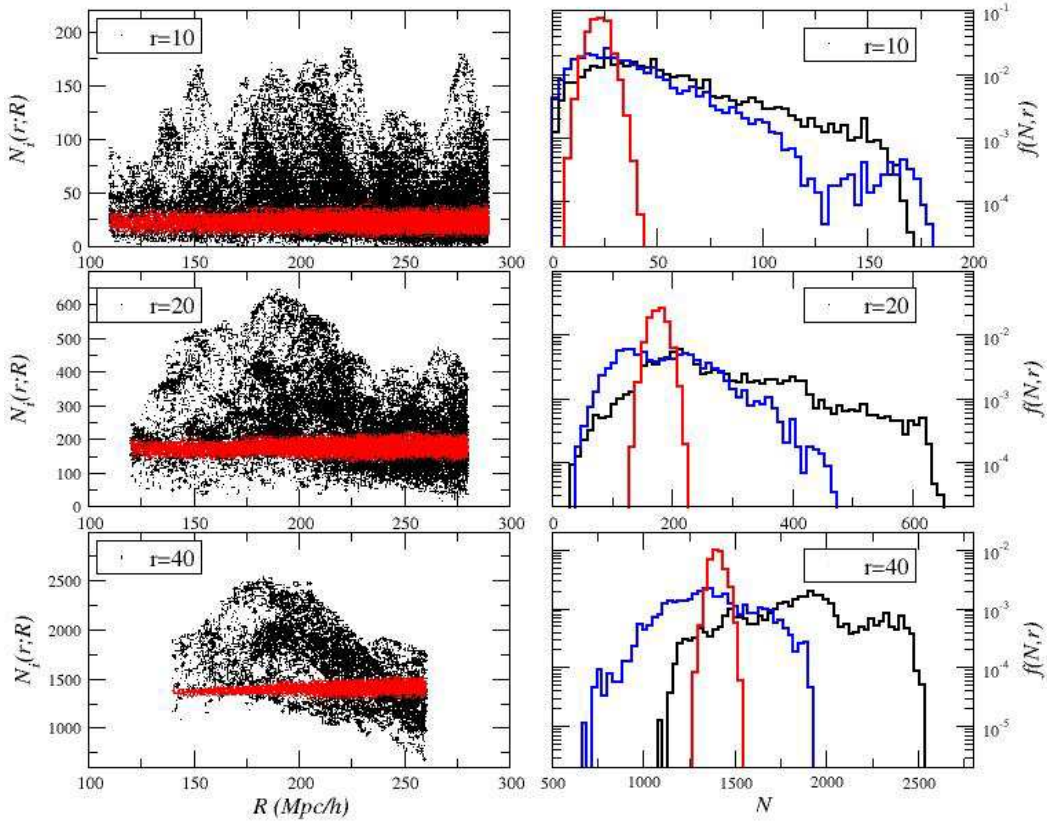} \caption{{\it
  Left panels}: From top to bottom the SL analysis a SDSS sample, with
  $r=10,20,40$ Mpc/h.  The red dots correspond to an uncorrelated
  distribution (Poisson) in the same volume and with same number of
  points.  {\it Right panels}: Probability density function of
  $N_i(r;R)$ in two non-overlapping sub-samples with equal volume
  (each half of the sample volume) at small and large $R$.  While for
  $r=10,20$ Mpc/h the PDF (nearby sub-sample black line and faraway
  sub-sample blue line) is reasonably statistically stable, for $r=40$
  Mpc/h there is a clear difference.  The red line corresponds to the
  Poisson distribution: a Gaussian function gives very good fits for
  all $r$ (adapted from \cite{sdss_aea}). }
\label{fig5} 
\end{figure}

In Fig.\ref{fig4} we show the three-dimensional representation of
$N(r;R_i)$ in a sample of the SDSS-DR6.  On the $X$ and $Y$ axis it is
reported the coordinate of the center of a sphere of radius $r=20$
Mpc/h (centered on a galaxy) and on the $Z$ axis the number of
galaxies inside it, normalized to its average value in this sample.
The mean thickness of this slice is about 50 Mpc/h.  Large
fluctuations in the density field traced by the SL analysis $N_i(r;R)$
are located in the correspondence of large scale structures.  The
information contained in the $N(r;R_i)$ data allow one to
quantitatively determine the properties of these structures in an
unambiguous way.
For instance we can determine the PDF of conditional fluctuations. An
example is reported in Fig.\ref{fig5}. In this sample, extracted from
the SDSS
\cite{sdss_aea}, the SL analysis detects large density fluctuations
without a clear radial-distance dependent trend.  Correspondingly the
PDF has a regular shape characterized by a peak with a long $N$ tail
and it is sufficiently statistically stable in different
non-overlapping sub-samples of equal volume.  This occurs except for
the largest sphere radii, i.e., for $r>$30 Mpc/h. The conclusion is
that conditional fluctuations in this sample are self-averaging for
$r<30$ Mpc/h, while they are not self-averaging for $r>30$ Mpc/h
because of the limited sample volumes.  These fluctuations determine
relative variations larger than unity in the estimation of the average
density, in the deepest samples, in spheres of radius $r=100$ Mpc/h.
The homogeneity scale must be $ > 100$ Mpc/h, the largest sphere
radius available in SDSS-DR6. The behavior of the PDF can a priori be
determined also by selection effects and not only by
fluctuations. However we tested that in the SDSS known selection
effects do not give a relevant contribution: we used volume-limited
samples and we made several tests to determine the importance of
cosmological corrections, such as different magnitude-distance
relations (i.e., different cosmological parameters), K-corrections and
evolutionary corrections \cite{sdss_aea}. It is interesting to note
that for $r<20$ Mpc/h the PDF of fluctuations displays a ``fat tail'',
and in this case an excellent fit is given by the Gumbel distribution
\cite{tibor}. 

Actually, in the largest samples of SDSS-DR7 we found \cite{tibor}
that over a large range of scales, both the average conditional
density and its variance show a nontrivial scaling behavior. The
average conditional density depends, for $10 \le r\le 80$ Mpc/h, only
weakly (logarithmically) on the system size.  Indeed, contrary to the
case of SDSS-DR6, in the larger sample volumes of DR7 fluctuations are
self-averaging up to $\sim 80$ Mpc/h.  Correspondingly, we find that
the density fluctuations follow the Gumbel distribution of extreme
value statistics. This distribution is clearly distinguishable from a
Gaussian distribution, which would arise for a homogeneous spatial
galaxy configuration. We concluded that there are similarities between
the galaxy distribution and critical systems of statistical physics,
which can display the same two features when correlations are
long-ranged.  Analogously in the 2dFGRS we found that the average
conditional density presents scaling behavior up to $\sim 30$ Mpc/h
and the PDF of conditional fluctuations show ``fat
tails'' \cite{2df_epl,2df_aea}.

Previous analyses of these galaxy catalogs, e.g., considered sample
averaged statistics without quantitatively testing whether a
significant bias could affect the results.  For instance the estimator
of the most commonly used statistics, the two-point correlation
function, can be written as \cite{book}
\be
\label{xi} 
\xi(r) +1 
\equiv \frac{\langle n(r) n(0)\rangle}{\langle n \rangle^2} 
= 
\lim_{V\rightarrow \infty} 
\frac{\overline{N(r,\Delta r)}} {V(r,\Delta r)} \cdot
\frac{V} {N} \;,  
\ee 
where in the second equality we have considered the finite sample
estimator (in the ensemble average sense the symbol $\overline{N}$
should be replaced by $\langle N \rangle$).  The first ratio in the
r.h.s. of Eq.\ref{xi} is the differential average conditional density,
i.e., the number of galaxies in shells of thickness $\Delta r$
averaged over the whole sample, divided by the volume $V(r,\Delta r)$
of the shell. The second ratio in the r.h.s. of Eq.\ref{xi} is the
average density estimated in a sample containing $N$ galaxies and with
volume $V$.  When measuring this function we implicitly assume, in a
given sample, that: (i) fluctuations are self-averaging in different
sub-volumes \cite{book} (ii) the linear dimension of the sample volume
is $V^{1/3} \gg \lambda_0$
\cite{book}, i.e., the distribution has reached homogeneity
inside the sample volume.  When the latter condition is not verified
the $\xi(r)$ analysis is biased by
systematic finite size effects even if 
fluctuations are self-averaging
\cite{book}.
These two assumptions can be tested directly by considering
conditional fluctuations properties, or indirectly by studying
finite-volume effects on $\xi(r)$ determinations. The result of the
$\xi(r)$ analysis is however in general not conclusive (see Sect 4.8
of \cite{sdss_aea}). For instance, in order to test how estimator bias
(e.g., the effect of the integral constraint) affects the results, one
can consider samples with different volumes and study whether there is
a convergence to a stable behavior, which was not found 
in the case of the LRG samples \cite{bao}. 

\begin{figure}
  \includegraphics[height=.5\textheight]{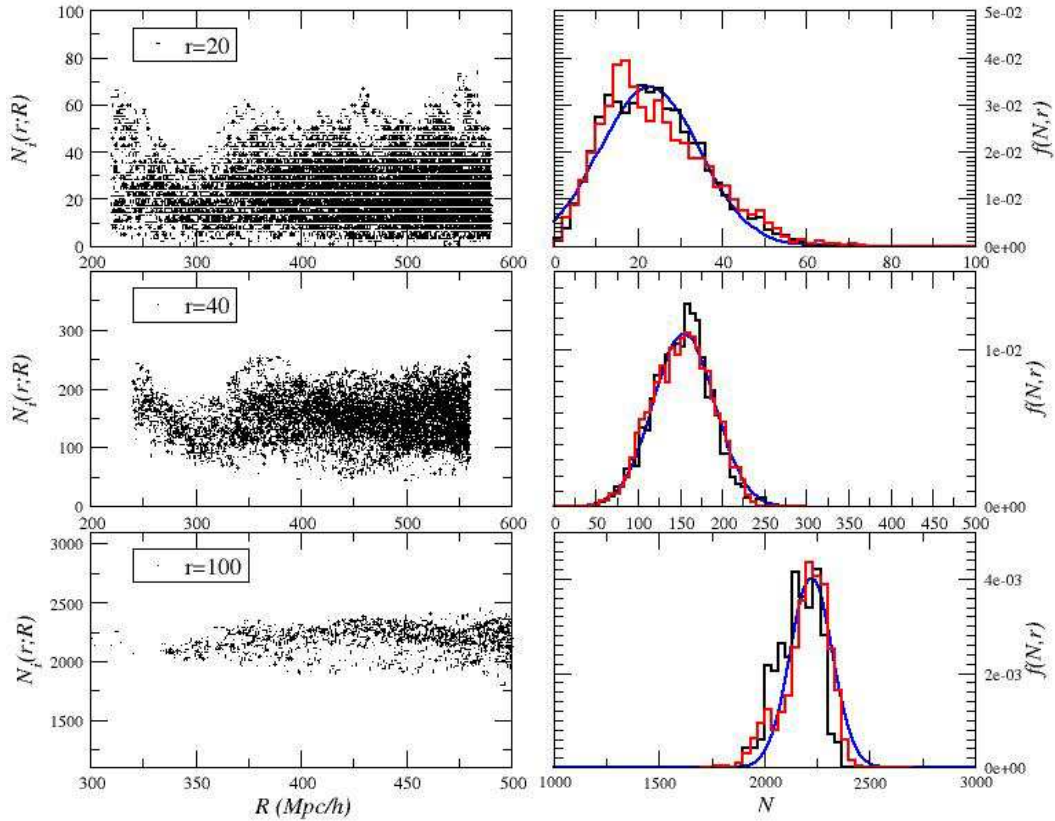} 
   \caption{ From
  top to bottom the SL analysis $N_i(r;R)$ for the mock sample VL2,
  with $r=20,40,100$ Mpc/h.  Although for $r=20$ Mpc/h fluctuations
  are still important, they rapidly become small for larger radii
  without any signature of a radial-distance dependent trend.  {\it
  Right panels}: Probability density function of $N_i(r;R)$ in two
  different non-overlapping sub-samples with equal volume (each half
  of the sample volume) at small (red line) and large (black line)
  $R$: the PDF is statistically stable for all $r$ and $R$.  The blue
  line corresponds to a Gaussian fit.}
\label{fig6} 
\end{figure}

Let us now discuss the case of a mock galaxy catalogs.  These are
generated from N-body simulations of standard cosmological
models \cite{springel05}, $N_i(r;R)$ does not show, for $r >10$ Mpc/h,
large fluctuations or systematic trends as a function of $R$
(Fig.\ref{fig6}).  
This is in agreement with the theoretical expectations
based on the linear growth of perturbations in an expanding universe
\cite{sdss_aea}. 
Because in these artificial catalogs fluctuations are small and
self-averaging, whole-sample averaged statistics are meaningful at all
scales.  Correspondingly the PDF rapidly converges to a Gaussian for
$r>10$ Mpc/h.  Thus mock catalogs, i.e. model predictions, are
self-averaging at all scales and uniform for $r>10$
Mpc/h \cite{sdss_aea}, and therefore the mock galaxy distribution is
qualitatively different from the real one. In agreement with our
results we note that \citet{einasto1} found that the fraction of very
luminous (massive) super-clusters in real samples extracted from
2dFGRS and from the SDSS (Data Release 4), is more than ten times
greater than in simulated samples. This again points toward a
non-trivial disagreement between the galaxy distribution and the mock
catalogs, stressing the fact that galaxy structures are more common in
observations than in the model.

\section{Conclusions}

We interpret the systematic differences found in the behavior of the
PDF of conditional fluctuations as due to a systematic effect in the
fact that sample volumes are not large enough for conditional
fluctuations, filtered at such large scales, to be self-averaging,
i.e.  to contain enough structures and voids of large size to allow a
reliable determination of average (conditional) quantities.  We
pointed out the problems related to the estimation of
amplitude of fluctuations
and correlation properties from statistical quantities which employ
the normalization to the estimation of the sample average. As long as
a distribution inside the given sample is not self-averaging, and thus
not homogeneous, the estimation of the two-point correlation function
is necessarily biased by strong finite size effects. Our results are
incompatible with homogeneity at scales smaller than $\sim 100$ Mpc/h.
While these results are at odds with LCDM predictions, they can be
compatible, at least theoretically, with a several recent observations
which also pose fundamental challenges to such a model.

For instance, Kashlinsky et al. \cite{Kashlinsky}, by studying the
fluctuations in the cosmic microwave background generated by the
scattering of the microwave photons by the hot X-ray emitting gas
inside clusters, have measured a coherent flow out to 300 Mpc/h with a
fairly high amplitude of 600-$10^3$ km/sec.  This is incompatible with
the standard LCDM model predictions. Indeed, on such large scales the
theoretical predictions are very simple, because in these models
gravitational clustering is still linear at those scales as density
fluctuations are small on large scales.  Similarly Watkins et
al. \cite{Watkins} estimated the bulk flow in all major peculiar
velocity surveys finding that the data suggest that the bulk flow
within a Gaussian window of radius 50 Mpc/h is 407 km/s.  They noticed
that this large-scale bulk motion indicates that there are significant
density fluctuations on very large scales. Indeed, a flow of this
amplitude on such a large scale is not expected in the LCDM model
cosmology, for which the predicted one-dimensional r.m.s. velocity is
about 110 km/s.  Thus the same problem for the model predictions we
found in the galaxy density field, are found also for the galaxy
velocity field. They may have the same origin, namely the fact that
there are galaxy structures which are too extended in space and have a
too large amplitude to be compatible with the predictions of the
standard model. In the case of the velocity field there is another
important element: the velocity field is generated by all the mass and
not only by the luminous component. Thus if the velocity field is so
high on such large scales, this may imply that this is generated by
the large scale inhomogeneities present in the overall mass
distribution, i.e. luminous plus dark. The fact that the whole mass
distribution is not homogeneous is compatible also with the results on
the matter density field derived by gravitational lens observations,
where very extended structures have been found \cite{massey}.  Thus
an important point which we aim to investigate in future works,
concerns the characterizing of the gravitational field generated by
galaxies.  When a distribution is inhomogeneous important
contributions to the gravitational force acting on a point can be due
to faraway sources \cite{force}.  The relation between the large scale
inhomogeneities of galaxy distribution and other observational data
should be examined in detail. For instance a statistically significant
anisotropy of the Hubble diagram at redshifts $z < 0.2$ was discovered
by \cite{sw08}. A local violation of statistical isotropy and
homogeneity, which may very well happen when matter distribution is
inhomogeneous
\cite{sl94,pwa}, 
can be related to such findings. although it is not excluded that a
systematic error in the observations or data analysis affect these
results.

Finally, it is worth mentioning that Joyce et al. \cite{pwa} pointed
out that an inhomogeneous distribution, as a fractal, does not
preclude the description of its gravitational dynamics in the
framework of the Friedmann-Robertson-Walker solutions to general
relativity. Indeed, the problem is often stated as being due to the
incompatibility of a fractal with the Cosmological Principle, where
this principle is identified with the requirement that the matter
distribution be isotropic and homogeneous. This identification is in
fact very misleading for a non-analytic and inhomogeneous structure
like a fractal, in which all points are equivalent statistically,
satisfying what has been called a Conditional Cosmological
Principle \cite{sl94,book}.  By treating the fractal as a perturbation
to an open cosmology in which the leading homogeneous component is the
cosmic background radiation one may get a simple explanation for the
supernovae data which indicate the absence of deceleration in the
expansion. This is indeed a very simplified theoretical model to interpret
the SN data and the large scale inhomogeneities in framework of the
Friedmann-Robertson-Walker metric.

\begin{theacknowledgments}
I am in debt with Yuri V. Baryshev and Nickolay L. Vasilyev for many
collaborations on this topic. I also thank Tibor Antal, Michael Joyce,
Andrea Gabrielli, Martin Lopez-Correidoira and Luciano Pietronero for
useful remarks and discussions. I am grateful to Michael Blanton and
David Hogg and for interesting comments.
\end{theacknowledgments}


\begin{thebibliography}{9}



\bibitem[Huchra et al., 1983]{cfa1}  
J. Huchra,  M. Davis,  Latham D., and Tonry J., 
{\it Astrophys.J.Suppl}, {\bf 52}, 89-119 (1983).



\bibitem[Falco et al., 1999]{cfa2} E.F. Falco, et al., 
 {\it Pub.Astron.Soc.Pac.}, {\bf 111},  438-452 (1999).


\bibitem[Giovanelli and Haynes, 1993]{pp}
R. Giovanelli, and M.P.  Haynes, {\it Astronom.J.}
{\bf 105}, 1271-1290 (1993). 


\bibitem[da Costa et al., 1988]{ssrs2} 
L. da Costa, et al., 
{\it Astrophys.J.}, {\bf 327}, 544-560 (1988). 

\bibitem[Shectman et al., 1996]{lcrs}  S. A. Shectman, et al., 
{\it Astrophys.J.}, {\bf 470}, 172-188 (1996).

\bibitem[York et al., 2000]{sdss}  D. York, et al., 
{\it Astronom.J.}  {\bf 120}, 1579-1587 (2000).

\bibitem[Colless et al., 2001]{colless01}
M.  Colless , et al., {\it Mon.Not.R.Acad.Soc}, {\bf 328}, 1039-1063
(2001).

\bibitem[de Lapparent Huchra and Geller, 1986]{dlhg85} 
V. de Lapparent, M.J.  Geller, J.P.  Huchra, 
{\it Astrophys.J.}, {\bf 302}, L1-L5 (1986).

\bibitem[Geller and Huchra, 1989]{gh89} M.J. Geller, J.P. Huchra 
{\it  Science}, {\bf 246}, 897-903  (1989).

\bibitem[Gott et al., 2005]{sloangreatwall} J.R. III Gott, , et al.,
{\it Astrophys.J.}, {\bf 624}, 463-484 (2005). 

\bibitem[Gabrielli et al., 2002]{glass} 
A. Gabrielli,  M.  Joyce, F. Sylos Labini, 
{\it Phys.Rev.}, {\bf D65}, 083523-1-083523-18 (2002).

\bibitem[Sylos Labini and Vasilyev, 2008] {cdm_theo} 
F. Sylos Labini, N.L.  Vasilyev,
{\it Astron.Astrophys.} {\bf 477}, 381-395 (2008).


\bibitem[Sylos Labini et al., 2009e]{bao} 
F.  Sylos Labini, N.L.  Vasilyev, Yu.V. Baryshev,
M. L\'opez-Corredoira, {\it Astron.Astrophys.}, {\bf 505}, 981-990
(2009)
 
\bibitem[Gabrielli et al., 2005]{book} 
 A. Gabrielli, F. Sylos Labini, M. Joyce and
L. Pietronero {\it ``Statistical physics for cosmic structures''}
Springer-Verlag (2005). 


\bibitem[Durrer et al., 2003]{bias} 
 R. Durrer, A. Gabrielli, M. Joyce, and  F. Sylos Labini,  
{\it Astrophys.J.}, {\bf 585}, L1-L4 (2003).


\bibitem[Sylos Labini et al., 2009a]{2df_epl} 
 F. Sylos Labini, N.  L. Vasilyev, Y.V. Baryshev, {\it
Europhys.Lett. }, {\bf 85 }, 29002-p1-29002-p6, (2009).


\bibitem[Sylos Labini et al., 2009b]{2df_aea} 
F. Sylos Labini, N.  L. Vasilyev, Y.V. Baryshev, {\it
Astron.Astrophys.}, {\bf 496}, 7-23 (2009).


\bibitem[Sylos Labini et al., 2009c]{sdss_epl} 
 F. Sylos Labini, N.  L. Vasilyev, L. Pietronero, Y.V. Baryshev, {\it
  Europhysics Letters}, {\bf 86}, 49001-p1-49001-p6 (2009).
  
\bibitem[Sylos Labini et al., 2009d]{sdss_aea} 
F. Sylos Labini, N.  L. Vasilyev, Y.V. Baryshev, {\it
Astron.Astrophys.}, in the press {\tt
    arXiv:0909.0132} (2009). 

\bibitem[Eisenstein et al., 2005]{lrg} 
D.J.  Eisenstein, {\it Astrophys.J.}, {\bf 633}, 560-574 (2005).

\bibitem[Cabr\'e and Gazta\~naga, 2008]{cabre} A. Cabr\'e, and 
E.  Gazta\~naga, 
{\it Mon.Not.R.Acad.Soc}, {\bf 396},  1119-1131 (2009).


\bibitem[Mart\'inez et al., 2009]{martinez} V.J Mart\'inez,  et al.
{\it Astrophys.J.}, {\bf 696}, L93-L97 (2009).


\bibitem[Kazin et al., 2009]{kazin} E. Kazin  et al., {\tt
    arXiv:0908.2598} (2009).


\bibitem[Loveday, 2004]{loveday}J. Loveday,  {\it Mon.Not.R.Acad.Soc}, {\bf
347}, 601-606 (2004).

\bibitem {busswell03} G.S. Busswell, et al., {\it Mon.Not.R.Acad.Soc} {\bf 354},
 991-1004 (2004).

\bibitem{frith03} W.J. Frith,  et al.,  {\it Mon.Not.R.Acad.Soc}, 
{\bf 345},   1049-1056 (2003).

\bibitem{frith06} W.J. Frith, et al., 
{\it   Mon.Not.R.Acad.Soc}, {\bf 371},  1601-1609 (2006). 

\bibitem{norbergxi01} E. Norberg, {\it et al.}, {\it Mon.Not.R.Acad.Soc},  
 {\bf 328}, 64-70 (2001). 

\bibitem{norbergxi02} E. Norberg, {\it et al.}, {\it Mon.Not.R.Acad.Soc}, 
 {\bf 332},  827-838 (2002).

\bibitem[Antal et al., 2009]{tibor} 
T. Antal,  F. Sylos Labini, N.  L. Vasilyev, Y.V. Baryshev
{\tt arXiv:0909.1507} (2009). 

\bibitem{springel05} V. Springel, et al.,  {\it
    Nature}, {\bf 435}, 629-636 (2005).
 
\bibitem[Einasto et al., 2006a]{einasto1} 
J. Einasto, et al., {\it Astron.Astrophys.}, {\bf 459}, L1-L4 (2006).

\bibitem[Kashlinsky et al., 2008]{Kashlinsky} A. Kashlinsky,  et al.,
{\it Astrophys.J.}, {\bf 686}, L49-L52 (2008).

\bibitem[Watkins et al., 2008]{Watkins} 
R. Watkins, H.A. Feldman, M.J. Hudson, 
{\it Mon.Not.R.Acad.Soc}, 
 {\bf 392}, 743-756 (2009). 



\bibitem[Massey et al., 2007]{massey} R. Massey, et al., 
{\it   Nature}, {\bf 445}, 286-290 (2007).


\bibitem{force}  A.  Gabrielli, F. Sylos Labini, \& S. Pellegrini, 
{\it Europhys.Lett.}, {\bf 46}, 127-133 (1999).


\bibitem[Schwarz and Weinhorst, 2007]{sw08} D.J. Schwarz, B. Weinhorst, 
{\it Astron.Astrophys.}, {\bf  474},717-729  (2007). 


\bibitem[Joyce et al., 2000]{pwa} 
  M. Joyce, P. W. Anderson, 
M. Montuori,   L. Pietronero and  F. Sylos Labini, 
{\it  Europhys.Letters}, {\bf 50}, 416-422 (2000).



\bibitem[Sylos Labini, 1994]{sl94}  F. Sylos Labini, 
{\it Astrophys.J.}, {\bf 433}, 464-467 (1994).




\end{thebibliography}
\end{document}